\documentclass[aps]{revtex4}
\usepackage{bm}
\usepackage{amsfonts,amssymb}
\usepackage{epsfig}

\begin{document}

\title{Control of chaos in Hamiltonian systems }
\author{G. Ciraolo}
\affiliation{Facolt\`a di Ingegneria, Universit\`a di Firenze,
via S. Marta, I-50129 Firenze, Italy, I.N.F.N. Sezione di Firenze and I.N.F.M. UdR Firenze}
\author{C. Chandre, R. Lima, M. Vittot}
\affiliation{CPT-CNRS, Luminy Case 907, F-13288 Marseille Cedex 9,
France}
\author{M. Pettini}
\affiliation{Istituto Nazionale di Astrofisica, Osservatorio
Astrofisico di Arcetri, Largo Enrico Fermi 5, I-50125 Firenze, Italy,
and I.N.F.M. UdR Firenze}

\begin{abstract}
We present a technique to control chaos in Hamiltonian systems which are close to integrable.
By adding a small and simple {\em control term} to the perturbation, the system becomes more regular than the original one. We
apply this technique to a forced pendulum model and show numerically that the control is able to drastically reduced chaos. 
\end{abstract}
\maketitle

\section{Introduction}
In this article, the problem we address is how to control chaos in 
Hamiltonian systems which are close to integrable.  
We consider the class of Hamiltonian systems that can be written
in the form $H=H_0+\epsilon V$ that is an integrable 
Hamiltonian $H_0$ (with action-angle variables)
plus a small perturbation $\epsilon V$.
\\ \indent The problem of control
in Hamiltonian systems is the following one: For the perturbed Hamiltonian
$H_0+\epsilon V$, the aim is to devise a control term $f$ such that
the dynamics of the controlled Hamiltonian $H_0+\epsilon V+f$
has more regular trajectories (e.g.~on invariant tori) or less diffusion
than the uncontrolled one. In practice, we do not require that the controlled Hamiltonian is integrable since it is a too strong requirement, but only that it has a more regular behavior than the original system.\\
Obviously $f=-\epsilon V$ is a solution since the resulting Hamiltonian is integrable. However, it is a useless solution
since the control is of the same order as the perturbation.  \\

 For practical purposes, the desired control term should be small
(with respect to the perturbation $\epsilon V$), localized in phase space
(meaning that the subset of phase space where $f$ is non-zero is finite
or small enough),
or $f$ should be of a specific shape (e.g. a sum of given Fourier modes, or with a certain regularity). Moreover, the control should be as simple as possible in order to be implemented in experiments. 
Therefore, the control appears to be a trade-off between the requirement on the reduction of chaos and the requirement on the simplicity of the control.\\
\\ \indent In this article, 
we provide an algorithm for finding a control term $f$
of order $O(\epsilon^2)$ such that $H=H_0+\epsilon V+ f$
is integrable. This control term is expressed as a series whose terms can be explicitly and easily computed by recursion.
It is shown on an example that truncations and approximations of this control term $f$ provides a simple and easy way to control the system.  \\

\section{Control theory of Hamiltonian systems.}
\label{sec:2}

In this section, we follow the exposition of control theory developed in Ref.~\cite{michel}. Let $\mathcal A$ be the algebra of real functions defined on
phase space. For $H\in \mathcal A$, let $\{H\}$ be the linear operator
acting on $\mathcal A$  such that
$$
\{H\}H^{\prime}=\{H,H^{\prime}\},
$$
for any $H^{\prime}\in \mathcal A$, where $\{\cdot~,\cdot\}$ is the Poisson bracket. The time-evolution of a function $V\in {\mathcal A}$ following the flow of $H$ is given by
$$
\frac{dV}{dt}=\{ H\} V,
$$  
which is formally solved as
$$
V(t)=e^{t\{H\}}V(0),
$$
if $H$ is time independent, and where
$$
e^{t\{H\}}=\sum_{n=0}^{\infty}\frac{t^n}{n!}\{H\}^n.
$$
Any element $V\in{\mathcal A}$ 
such that \( \{{H}\}{V} =0 \), is constant under the flow of \( {H} \),
i.e.
$$
\forall t\in {\mathbb R}, \qquad e^{{t} \{{H}\}} {V} = {V}.
$$
Let us now fix a Hamiltonian $H_0\in{\mathcal A}$.
The vector space \( \mathrm{Ker} \{{H_0}\} \) is the set of constants
of motion and it is a sub-algebra of $\mathcal A$. The operator
\(\{{H_0}\} \) is not invertible since a derivation has always 
a non-trivial kernel. For
instance \( \{ {H_0} \} ({H_0}^\alpha) = 0 \) for any $\alpha$ such
that \( {H_0}^\alpha \in {\mathcal A} \). 
Hence we consider a pseudo-inverse of \( \{{H_0}\} \).
We define a linear operator $\Gamma$ on $\mathcal A$ such that
\begin{equation}
\{{H_0}\}^{2}\ \Gamma = \{{H_0}\},
\label{gamma}
\end{equation}
i.e.
$$
\forall V\in {\mathcal A}, \qquad \{H_0,\{H_0,\Gamma V\}\}=\{H_0,V\}.
$$
If the operator $\Gamma$ exists, it is not unique in general. Any other choice
$\Gamma^{\prime}$ satisfies $\rm{Rg}(\Gamma^{\prime}-\Gamma)\in \rm{Ker}(\{H_0\}^2)$.
\\ \indent We define the {\em non-resonant} operator $\mathcal N$ and the 
{\em resonant} operator $\mathcal R$ as
\begin{eqnarray*}
&& {\mathcal N} = \{H_0\}\Gamma,\\
&& {\mathcal R} = 1-{\mathcal N},
\end{eqnarray*}
where the operator $1$ is the identity in the
algebra of linear operators acting on \( \mathcal {A} \). 
We notice that Eq.(\ref{gamma}) becomes
$$
\{{H_0}\} \mathcal R = 0,
$$
which means that the range \( {\rm Rg} \mathcal R \) of the operator \( \mathcal R \) is
included in \( {\rm Ker} \{{H_0}\} \).
A consequence
is that any element ${\mathcal R} V$ is constant under the flow of $H_0$, i.e. 
$e^{t\{H_0\}}{\mathcal R}V={\mathcal R}V$. We notice that when $\{H_0\}$
and $\Gamma$ commute, $\mathcal R$ and $\mathcal N$ are projectors, i.e.
$\mathcal R^2=\mathcal R$ and $\mathcal N^2=\mathcal N$. Moreover, in this case we have ${\rm Rg} \mathcal R = {\rm Ker} \{{H_0}\} $, i.e.\ the constant of motion are the elements $\mathcal{R}V$ where $V\in \mathcal{A}$.
\\ \indent Let us now assume that $H_0$ is integrable with action-angle variables 
$({\bf A},\bm{\varphi})\in B\times {\mathbb T}^n $ where $B$ is an open set
of $\mathbb R^n$ and ${\mathbb T}^n$ is the $n$-dimensional torus, so that
$H_0=H_0({\bf A})$ and the Poisson bracket $\{H,H^{\prime}\}$ between two Hamiltonians is 
$$
\{H,H^{\prime}\}=\frac{\partial H}{\partial{\bf A}}\cdot
\frac{\partial H^{\prime}}{\partial{\bm{\varphi}}}-
\frac{\partial H}{\partial{\bm{\varphi}}}\cdot
\frac{\partial H^{\prime}}{\partial{\bf A}}.
$$
The operator $\{H_0\}$ acts on $V$ given by
$$
V=\sum_{{\bf k}\in \mathbb Z^n}V_{\bf k}({\bf A})e^{i{\bf k}\cdot{\bm\varphi}},
$$
as
$$
\{H_0\}V({\bf A},\bm{\varphi})=\sum_{\bf k}i{\bm \omega}({\bf A})\cdot{\bf k}~V_{\bf k}({\bf A})e^{i{\bf k}\cdot\bm\varphi},
$$
where the frequency vector is given by
$$
{\bm \omega}({\bf A})=\frac{\partial H_0}{\partial{\bf A}}. 
$$
A possible choice of $\Gamma$ is
$$
\Gamma V({\bf A},\bm{\varphi})=
\sum_{{\bf k}\in{\mathbb Z^n}\atop{\omega({\bf A})\cdot{\bf k}\neq0}}
\frac{V_{\bf k}({\bf A})}
{i{\bm \omega}({\bf A}) \cdot{\bf k}}~~e^{i{\bf k}\cdot{\bm\varphi}}.
$$
We notice that this choice of $\Gamma$ commutes with $\{H_0\}$.
\\ \indent For a given $V\in{\mathcal A}$, ${\mathcal R} V$ is the resonant 
part of $V$ and ${\mathcal N} V$ is the non-resonant part:
\begin{eqnarray}
&&{\mathcal R}V=\sum_{\bf k~}
V_{\bf k}({\bf A})\chi(\bm\omega({\bf A})\cdot{\bf k}=0)e^{i{\bf k}\cdot{\bm\varphi}},\label{eqn:RV}\\ 
&&{\mathcal N}V=\sum_{\bf k~}
V_{\bf k}({\bf A})\chi(\bm\omega({\bf A})\cdot{\bf k}\neq0)e^{i{\bf k}\cdot{\bm\varphi}},
\end{eqnarray}
where $\chi(\alpha)$ vanishes  when proposition $\alpha$ is wrong and it is equal to $1$
when  $\alpha$ is true.\\

From these operators defined for the integrable part $H_0$, we construct a control term for the perturbed Hamiltonian $H_0+V$ where $V\in {\mathcal A}$, i.e.\ we construct $f$ such that $H_0+V+f$
is canonically conjugate to $H_0+\mathcal R V$.\\

\noindent {\em Proposition 1: } ~~For $V \in {\mathcal A}$ and $\Gamma$ constructed
from $H_0$, we have the following equation
\begin{equation}
e^{\{\Gamma V\}}(H_0+V+f)=H_0+{\mathcal R} V,
\label{prop1}
\end{equation}
where
\begin{equation}
f(V)=e^{-\{\Gamma V\}}{\mathcal R}V+\frac{1-e^{-\{\Gamma
V\}}}{\{\Gamma V\}} {\mathcal N} V -V.
\end{equation}

We notice that the operator $(1-e^{-\{\Gamma V\}})/\{\Gamma V\}$
is well defined by the expansion
$$
\frac{1-e^{-\{\Gamma V\}}}{\{\Gamma V\}}=
\sum_{n=0}^{\infty}\frac{(-1)^n}{(n+1)!}\{\Gamma V\}^n.
$$
We can expand the control term in power series as
\begin{equation}
f(V)=\sum_{n=1}^{\infty}\frac{(-1)^n}{(n+1)!}\{\Gamma V\}^n
(n{\mathcal R}+1)V.
\label{expansion_f}
\end{equation}
\\
We notice that
if $V$ is of order $\epsilon$, $f(V)$ is of order $\epsilon^2$.\\
\indent Proposition 1 tells that the addition of a well chosen 
control term $f$ makes the Hamiltonian canonically
conjugate to $H_0+{\mathcal R} V$. 

\noindent {\em Proposition 2 :} ~~The flow of $H_0+V+f$ is conjugate to the flow of $H_0+{\mathcal R}V$:
$$
\forall {t} \in {\mathbb R} , \qquad e^{{t} \{{H_0} + {V} + {f}\}} =
e^{-\{\Gamma {V}\}} ~ e^{{t} \{{H_0}\}}~ e^{{t} \{\mathcal R {V}\}}
~ e^{\{\Gamma {V}\}}. 
$$

The remarkable fact is that
the flow of ${\mathcal R}V$ commutes with the one of $H_0$, since
$\{H_0\}{\mathcal R} = 0$. This allows the splitting of the flow of 
$H_0 +{\mathcal R} V$ into a product. \\
 
We recall that
$H_0$ is {\em non-resonant} iff
$$
\forall {\bf A}\in B, ~~  \chi \left(\omega({\bf A})\cdot{\bf k}=0\right)=\chi ( {\bf k=0} ).
$$ 
If $H_0$ is non-resonant then with the addition of a
control term $f$, the Hamiltonian $H_0+V+f$ is 
canonically conjugate to the integrable Hamiltonian $H_0+{\mathcal R} V$
since ${\mathcal R} V$ is only a function of the 
actions [see Eq.~(\ref{eqn:RV})].
\\ \indent If $H_0$ is resonant and ${\mathcal R} V=0$, the controlled
Hamiltonian $H=H_0+V+f$ is conjugate to $H_0$.
\\ In the case ${\mathcal R} V=0$, the series (\ref{expansion_f})
which gives the expansion of the control term $f$,
can be written as
\begin{equation}
f(V)=\sum_{s=2}^{\infty}f_s,
\label{exp_f_rv_0}
\end{equation}
where $f_s$ is of order $\epsilon^s$ and given by the
recursion formula
\begin{equation}
f_s=-\frac{1}{s}\{\Gamma V,f_{s-1}\},
\label{recursion}
\end{equation}
where $f_1=V$.

\noindent {\em Remark :} A similar approach of control has been developed by G. Gallavotti in Refs.~\cite{gallavotti2,gallavotti1,gentile}. The idea is to find a control term (named {\it counterterm}) only depending on the actions, i.e.\ to find $N$ such that
$$
H({\bf A},{\bm \varphi})=H_0({\bf A})+V({\bf A},{\bm \varphi})-N({\bf A})
$$
is integrable. For isochronous systems, that is
$$
H_0({\bf A})={\bm \omega}\cdot {\bf A}
$$
it is shown that if the frequency vector satisfies a Diophantine condition and if the perturbation is sufficiently small and smooth, such a control term exists, and that an algorithm to compute it by recursion is provided by the proof. We notice that the resulting control term $N$ is of the same order as the perturbation.

\section{Application to a forced pendulum model}

\begin{figure}[ht]
\unitlength 1cm
\begin{picture}(10,9)(0,0)
\put(0,0){\epsfig{file=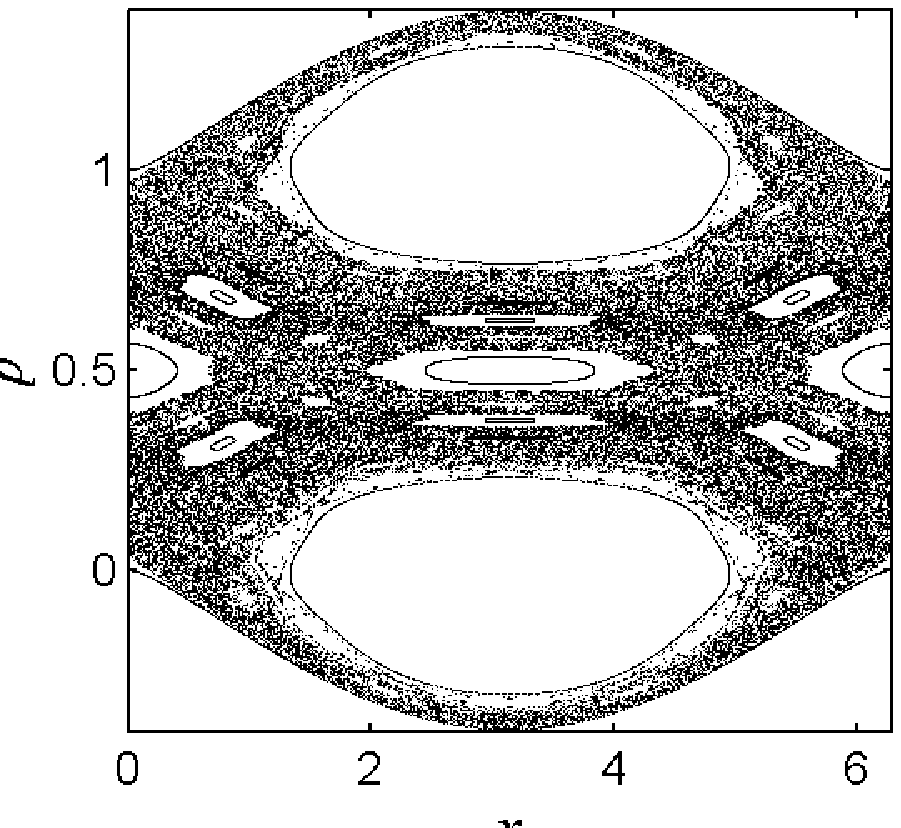,width=10cm,height=9cm}}
\put(4.5,5.8){\epsfig{file=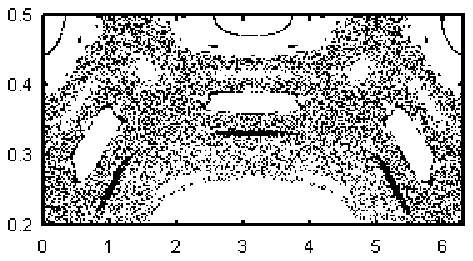,width=5.3cm,height=3cm}}
\end{picture}
\caption{Poincar\'e surface of section of Hamiltonian (\ref{eqn:fp}) with $\varepsilon=0.034$.}
\label{fig1}
\end{figure}

\begin{figure}[ht]
\unitlength 1cm
\begin{picture}(10,9)(0,0)
\put(0,0){\epsfig{file=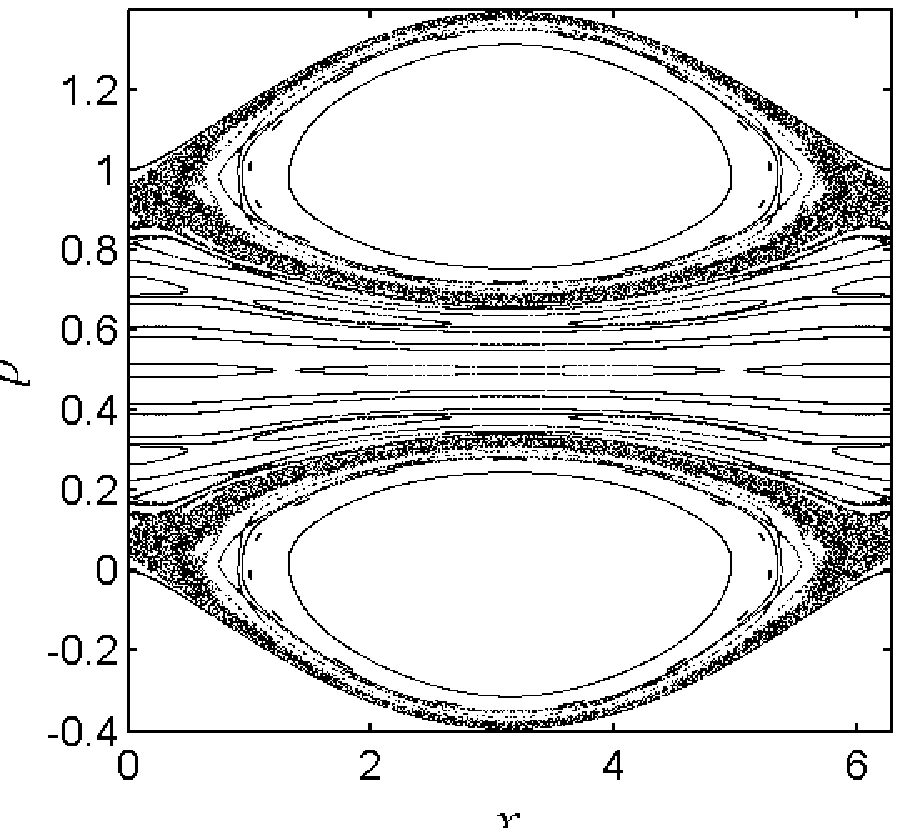,width=10cm,height=9cm}}
\put(4.5,5.8){\epsfig{file=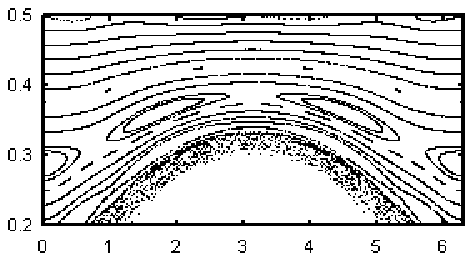,width=5.3cm,height=3cm}}
\end{picture}
\caption{Poincar\'e surface of section of Hamiltonian (\ref{eqn:fp}) with the approximate control term~(\ref{eqn:f2fpa}) with $\varepsilon=0.034$.}
\label{fig2}
\end{figure}

\begin{figure}[ht]
\unitlength 0.7cm
\begin{picture}(17,8)(0,0)
\put(0,0){\epsfig{file=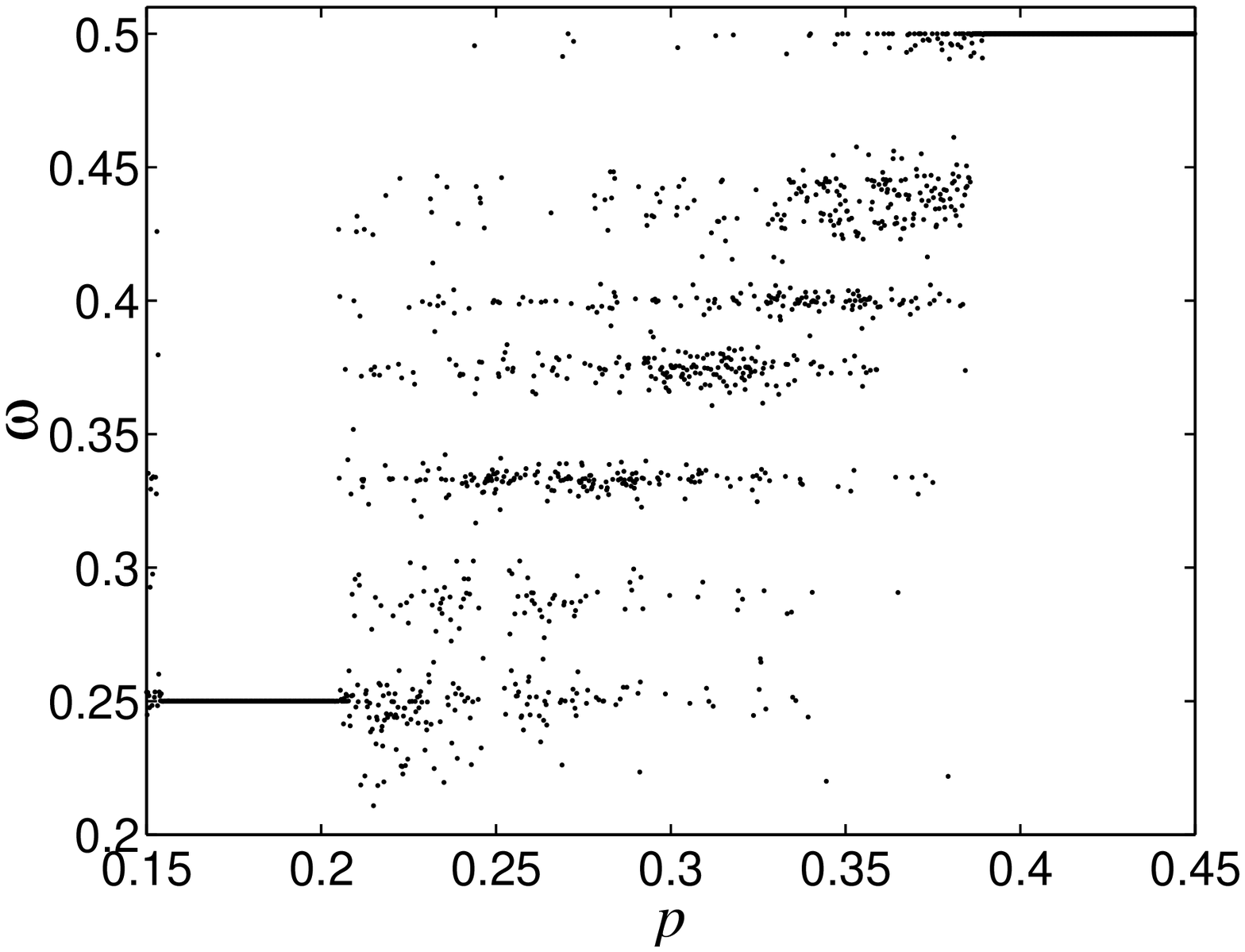,width=5.7cm,height=5.2cm}}
\put(8.1,0){\epsfig{file=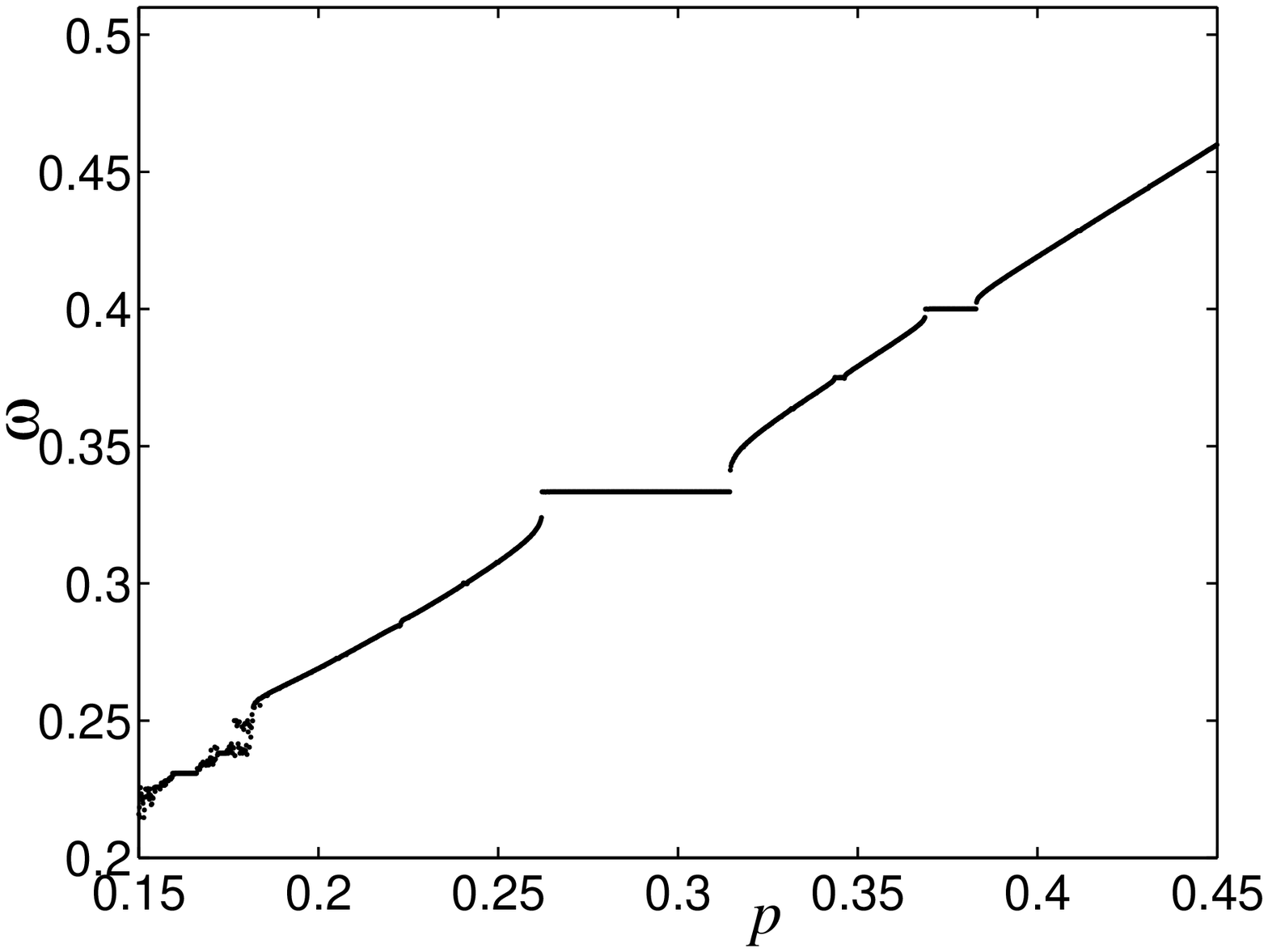,width=5.7cm,height=5.2cm}}
\end{picture}
\caption{Frequency analysis of Hamiltonian (\ref{eqn:fp}) with $\varepsilon=0.034$ $(a)$without control term and $(b)$ with control term~(\ref{eqn:f2fpa}). The fundamental frequency $\omega(p)$ is plotted versus $p$ for $p\in [0.15, 0.45]$ for the trajectories with initial conditions $(x=0,p)$. }
\label{figfma1}
\end{figure}

We consider the following model with 1.5 degrees of freedom
\begin{equation}
\label{eqn:fp}
H(p,x,t)=\frac{1}{2}p^2+\varepsilon \left[ \cos x+\cos(x-t)\right].
\end{equation}
Figure~\ref{fig1} depicts a Poincar\'e section of Hamiltonian~(\ref{eqn:fp}) for $\varepsilon=0.034$. We notice that for $\varepsilon\geq 0.02759$ there are no longer any KAM tori~\cite{chanPR}.
In order to apply the control theory described above, we need to put Hamiltonian~(\ref{eqn:fp}) in an autonomous form. We consider that $t$ is an additional angle whose conjugate action is $E$. The Hamiltonian becomes autonomous
\begin{equation}
\label{eqn:fpa}
H(p,x,E,t)=\frac{1}{2}p^2+E+\varepsilon \left[ \cos x+\cos(x-t)\right],
\end{equation}
where the actions are ${\bf A}=(p,E)$ and the angles are $\bm{\varphi}=(x,t)$.
The unperturbed Hamiltonian that will be used to construct the operators $\Gamma$, $\mathcal{R}$ and $\mathcal{N}$ is
\begin{equation}
\label{eqn:H0}
H_0(p,E)=\frac{1}{2}p^2+E.
\end{equation}
The action of $\{H_0\}$, $\Gamma$, $\mathcal{R}$ and $\mathcal{N}$ on functions $V\in {\mathcal A}$ given by
$$
V(p,x,E,t)=\sum_{k_1,k_2\in \mathbb{Z}} V_{k_1,k_2}(p,E)e^{i(k_1x+k_2t)},
$$
is
\begin{eqnarray*}
&& \{H_0\} V=\sum_{k_1,k_2\in \mathbb{Z}} i(pk_1+k_2) V_{k_1,k_2}(p,E)e^{i(k_1x+k_2t)},\\
&& \Gamma V=\sum_{k_1,k_2\in \mathbb{Z}} \frac{\chi(pk_1+k_2\not=0)}{i(pk_1+k_2)}V_{k_1,k_2}(p,E)e^{i(k_1x+k_2t)},\\
&& {\mathcal R} V=\sum_{k_1,k_2\in \mathbb{Z}} \chi(pk_1+k_2=0) V_{k_1,k_2}(p,E)e^{i(k_1x+k_2t)},\\
&& {\mathcal N} V=\sum_{k_1,k_2\in \mathbb{Z}} \chi(pk_1+k_2\not=0) V_{k_1,k_2}(p,E)e^{i(k_1x+k_2t)}.
\end{eqnarray*}
The action of these operators on $V(x,t)=\varepsilon \left[ \cos x+\cos (x-t)\right]$ is
\begin{eqnarray*}
&& \{H_0\} V= -\varepsilon \left[ p\sin x+(p-1)\sin(x-t)\right],\\
&& \Gamma V=\varepsilon \left[ \frac{1}{p}\sin x+\frac{1}{p-1}\sin(x-t)\right],\\
&& {\mathcal R} V=0,\\
&& {\mathcal N} V=V=\varepsilon \left[ \cos x+\cos (x-t)\right],
\end{eqnarray*}
for $p\not= 0,1$.
Since $\mathcal{R} V=0$, the control term is given by Eq.~(\ref{exp_f_rv_0}) and the terms in the series are computed by recursion. In particular, the first term $f_2$ is given by
$$
f_2=-\frac{1}{2}\{ \Gamma V, V\}=-\frac{1}{2}\frac{\partial \Gamma V}{\partial p } \frac{\partial V }{\partial x},
$$
since $V$ is independent of $p$. The explicit expression of $f_2$ is given by
\begin{eqnarray}
f_2(p,x,t)=&&\frac{\varepsilon^2}{4}\left( \frac{1}{p^2}\cos 2x +\frac{1}{(p-1)^2}\cos 2(x-t)\right)\nonumber \\
&& - \frac{\varepsilon^2}{4} \left( \frac{1}{p^2}+\frac{1}{(p-1)^2}\right) \left(1+\cos t -\cos (2x-t)\right).\label{eqn:f2fp}
\end{eqnarray}
The main purpose of the control is to have a control term which is as simple as possible in order to be implemented in the experiment.\\
A possible simplification is to consider the region in between the two primary resonances located around $p=0$ and $p=1$. We develop the approximate control term around $p=1/2$. It thus becomes
$$
f_2(p,x,t)=\varepsilon^2\left( \cos 2x +\cos 2(x-t) +2\cos (2x-t)\right),
$$
up to a function that only depends on time.
Furthermore, we only keep the main Fourier mode of this control term. In the region $p=1/2$ the two terms $\cos 2x $ and $\cos 2(x-t)$ which are associated with resonances approximately located at $p=0$ and $p=1$  have a much smaller influence than the term $\cos (2x-t)$ which is associated with a resonance located at $p=1/2$. The control term we study is thus
\begin{equation}
\label{eqn:f2fpa}
f_2(p,x,t)=2\varepsilon^2\cos (2x-t).
\end{equation}
Figure~\ref{fig2} depicts a Poincar\'e section of the Hamiltonian~(\ref{eqn:fp}) with the approximate control term~(\ref{eqn:f2fpa}) for the same value of $\varepsilon$ used in Fig.~\ref{fig1}, i.e.\ $\varepsilon=0.034$. It clearly shows that a lot of KAM tori are created with the addition of the control term. 
We notice that the perturbation has a norm (defined as the maximum of its amplitude) of $6.8\times 10^{-2}$ whereas the control term has a norm of $2.3\times 10^{-4}$ for $\varepsilon=0.034$. The control term is small (less than 4\% ) compared to the perturbation . \\
We can decrease the amplitude of the control term by considering
$$
f_2(p,x,t)=\alpha \varepsilon^2\cos (2x-t).
$$
where $\alpha \leq 2$ and still get a lot of KAM tori in the vicinity of the region $p=1/2$. For instance, the control is still effective with $\alpha=0.8$.
Using the renormalization-group transformation~\cite{chanPR}, we have looked at the domain of existence of the golden-mean KAM torus which is the rotational invariant torus with frequency $(3-\sqrt{5})/2$. This rotational invariant torus only exists in the domain
$$
\alpha \in [0.778, 3.173].
$$
Therefore, in this domain of $\alpha$ there are barriers in phase space that prevent diffusion of trajectories. The control is optimal for $\alpha=2$. Increasing the amplitude of the control does not improve the control which means that the control we devised is not "brute force". This control is robust since it is effective away from its reference value $\alpha=2$. There is also the possibility of reducing the control (by a factor larger than 2) and still get a significant effect of the control.   \\
In order to see more precisely the effect of the control term, we have applied Laskar's Frequency Map  Analysis~\cite{laskar} to this model. For initial conditions $(x,p)$ where $x=0$ and $p\in [0.15, 0.5]$, we compute the main (rotational) frequency $\omega (p)$ of the trajectory integrated from 0 to $T=2.5\times 10^3$ following Laskar's procedure. Figure~\ref{figfma1} depicts the frequency $\omega$ as a function of $p$ for Hamiltonian~(\ref{eqn:fp}) with $\varepsilon=0.034$ with and without the addition of the approximate control term~(\ref{eqn:f2fpa}). An apparent continuous variation of the frequency is associated with a region with KAM tori. A flat variation is associated with a resonant island. A scatter of points is associated with a chaotic region.\\
Without control term, the region located between the stable regions of the resonances 1:4 and 1:2 is chaotic and does not contain any KAM tori since the frequency map does not show continuous regions. With control term, we clearly see that the frequency curve appears to be continuous and hence the region contains a lot of KAM tori.

\section{Conclusion} We have shown on a simple example the explicit construction of a control term. This term can be drastically simplified and still is able to reduce chaos in the system. The idea is to have a control term as simple as possible in order to be implemented in experiment. We notice that this approach has been applied to a model of transport for the ${\bf E}\times {\bf B}$ drift~\cite{guido1,guido2}. It has been shown that the control drastically reduces the chaotic transport in this model.

\acknowledgments 
We acknowledge useful discussions with G.\ Gallavotti and J. Laskar.

\end{document}